# Degenerate merging BICs in resonant metasurfaces

YIXIAO GAO,[1,5,6] JUNYANG GE,[1,5] ZHAOFENG GU,[1,5] LEI XU,[2] XIANG SHEN,[1,4] AND LUJUN HUANG[3,7]

[1]*Laboratory of Infrared Materials and Devices, Zhejiang Key Laboratory of Photoelectric Detection Materials and Devices, Research Institute of Advanced Technologies, Ningbo University, Ningbo, Zhejiang 315211, China*
[2]*Advanced Optics & Photonics Laboratory, Department of Engineering, School of Science and Technology, Nottingham Trent University, Nottingham NG11 8NS, United Kingdom*
[3]*School of Physics and Electronic Science, East China Normal University, Shanghai 200241, China*
[4]*Ningbo Institute of Oceanography, Ningbo 315832, China*
[5]*Equally contributed*
[6]*gaoyixiao@nbu.edu.cn*
[7]*ljhuang@phy.ecnu.edu.cn*

**Resonant metasurfaces driven by bound states in the continuum (BIC) offer an intriguing approach to engineer high-Q resonances. Merging multiple BICs in the momentum space could further enhance the Q-factor as well as its robustness to fabrication imperfections. Here, we report doubly-degenerate guided mode resonances (GMR) in a resonant metasurface, whose radiation losses could be totally suppressed due to merging BICs. We show that the GMRs and their associated accidental BICs can be evolved into degenerate merging BICs by parametric tuning of the metasurface. Significantly, these two GMRs share the same critical parameter (i.e. lattice constants or thickness) that the merging BICs occur. Interestingly, thanks to the degenerate property of two GMRs, a larger (smaller) period will split one of merging BICs into eight accidental BICs at off-Γ point, but annihilate the other. Such exotic phenomenon can be well explained from the interaction of GMRs and background Fabry-Perot resonances. Our result provides new strategies to engineering high-Q resonances in resonant metasurfaces for light-matter interaction.**

Flat optics based on metasurfaces become the major driving force in miniaturizing conventional bulky optical elements for light field control. When long-range interaction among neighboring unit cells becomes pronounced, metasurfaces exhibit resonant behavior with sharp spectral features [1–3], which have been widely explored in biosensing [4], fluorescence control [5], microlasers [6], and nonlinear applications [7,8]. Various physical mechanisms are explored to construct high-$Q$ resonant metasurfaces, including guided mode resonance (GMR) [9,10], surface lattice resonance [11], as well as the newly emerged concept of bound states in the continuum (BICs) [12]. The $Q$-factors of the former two types of resonances are influenced by the scatterer size in the array, and high-$Q$ resonances usually requires small scatterer size, posing fabrication challenge. BIC is a class of localized state embedded within the frequencies of a continuum band but decouple with free space radiations [12]. Theoretically, the resonant metasurfaces driven by the physics of BICs could give rise to an infinite $Q$-factor, and practical applications require these resonant metasurfaces working in quasi-BIC states with finite $Q$ factors. For achieving high-$Q$ resonance, symmetry-protected BICs typically requires weak geometric perturbations [13,14], while fabrication imperfections can significantly deteriorate these Q factors [14,15]. Integrating tunable materials into resonant metasurfaces may provide a potential solution, though, it requires additional fabrication steps [16]. Symmetry perturbations near weak field region within the unit cell could open the radiation channel while preserving high-$Q$ resonance robustness against fabrication errors [17,18], and Huang et al. realized a high-quality resonant metasurface with a $Q$ factor of 36694 based on such principle [19]. On the other hand, practically fabricated resonant metasurface are finite in size, $Q$-factor of the resonance is usually lower than the theoretical calculation where the metasurface is assumed as infinitely extended. Therefore, a general quest to boost the $Q$-factors of quasi-BICs is desired for compact nanophotonic structures.

Merging BICs offers an promising method to enhance the $Q$-factor of resonant metasurfaces [20–23]. Symmetry-protected (SP) BICs usually associate with several accidental BICs at off-Γ points, originated from destructive interference of multipole fields [24]. Typically, the position of accidental BICs in the momentum space is dependent on certain metasurface parameters (e.g. period). By continuously tuning the metasurface parameter, accidental BICs could gradually approach towards and merges at the Γ-point, leading to greatly improved $Q$-factors near Γ point. For example, the $Q$-factor of a single BIC increase quadratically following the scaling law as $Q \propto 1/k^2$, where $k$ is the deviated wavenumber in the momentum space from the isolated BIC. While Ref. [20] reports the $Q$-factor scaling law becomes $Q \propto 1/k^6$ after merging nine accidental BICs at Γ point, leading to a drastically enhanced $Q$-factor being orders of magnitude higher than those in an isolated-BIC [20,21], especially for finite-sized resonant metasurfaces [25]. Merging BICs are also more robust to structural defects compared to isolated BICs [20]. Previous works mainly focus on the merging process towards a single non-degenerate resonance [20,21,23], while merging accidental BICs with degenerate resonances are rarely studied.

In this paper, we demonstrate degenerate merging BICs in resonant metasurfaces. We focus on a doubly-degenerate GMRs in a square lattice of dielectric pillars embedded in a homogeneous medium, and the radiation loss of this mode could be totally suppressed at a critical lattice constant, leading to the formation of degenerate merging BIC. We show that there are eight accidental BICs at off-Γ points for each GMRs. They will merge into a single one with tuning the structural parameters (i.e. period or thickness). We also investigate resonant characteristics of degenerate merging BICs including the $Q$-factor scaling law ($Q \propto 1/k^4$), the geometric condition where degenerate merging BIC could occur. Our findings would provide new insights of merging BIC behavior within degenerate modes, which is also of practical interest for planar photonic applications requiring high quality resonances.

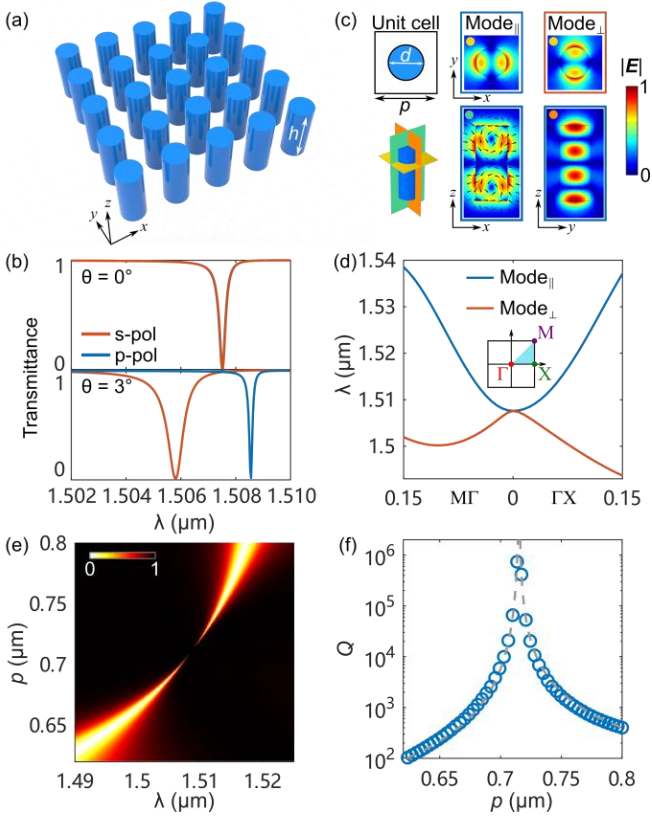

Fig. 1. (a) Schematic for the resonant metasurface consisting of a square lattice of dielectric pillars. (b) Transmission spectrum of the studied metasurface under normal and oblique incident plane wave. Lattice constant $p$ = 700 nm. (c) The electric field profiles of the two degenerate modes: Mode$_\parallel$ and Mode$_\perp$. (d) Dispersion relation of Mode$_\parallel$ and Mode$_\perp$ along the MΓ and ΓX directions. Inset shows the first Brillouin zone of the square lattice. (e) Evolution of transmission spectra of the resonant metasurface under normal excitation with varying lattice constant $p$. (f) $Q$-factor of the resonance as a function of $p$.

Figure 1(a) shows the schematic of the studied resonant metasurface consisting of a square lattice of dielectric pillars. Each pillar has a diameter of $d$ and height of $h$. The lattice constant is $p$ along both $x$ and $y$ directions. The dielectric pillars are made of silicon with a refractive index of 3.5, embedded in a homogenous medium with a refractive index of 1. Here we set $d$ = 400 nm, $h$ = 950 nm, and $p$ = 700 nm. Although the aspect ratio of the structure is larger than 2, previous studies of all-dielectric metalens have demonstrated the fabrication feasibility with the standard nanofabrication technology due to the mature fabrication process of silicon [26]. In such configuration, a normally-incident plane wave shine on an extended metasurface would excite a sharp resonance at 1507.5 nm in the transmission spectrum, as plotted in the upper panel of Fig. 1(b). Owing to its $C_{4v}$ symmetry of the lattice structure, the spectral response is independent on the incident polarizations. An eigenfrequency calculation of the resonant modes in the unit cell reveals that the transmission dip corresponds to the excitation of a doubly-degenerate GMR mode, as depicted in Fig. 1(c). The vectorial electric field, as shown by the dark arrows, suggests the resonance field is originated from a pair of vertically stacked magnetic dipoles, this resonance is also utilized for local phase modulation in dielectric metalenses [27]. When the metasurface is excited by an oblique incidence, corresponding to the presence of an in-plane wavevector, the mode degeneracy is lifted, manifested by a splitting resonant wavelength upon increasing $k_\parallel$, as depicted in the lower panel of Fig. 1(b). Figure 1(d) shows the band structure of two GMR modes. Here Γ, X, and M represent the high-symmetry points of the first Brillouin zone for the square lattice, as denoted in the inset of Fig. 1(d). In what follows, we refer to the mode with its electric field circulating parallel to $k_\parallel$ (or its magnetic dipole moment orthogonal to $k_\parallel$) as Mode$_\parallel$, while the other resonance with its electric field circulating perpendicular to $k_\parallel$ (or its magnetic dipole moment parallel to $k_\parallel$) is named as Mode$_\perp$, as depicted in Fig. 1(c).

Figure 1(e) shows the transmission spectra of a normal incident plane wave under varying $p$. The transmittance is obtained by a rigorous coupled wave analysis (RCWA) [28]. With $p$ increasing, the resonant wavelength shows a redshift. However, the linewidth of the transmission dip demonstrates a nontrivial characteristic: a vanishing linewidth could be observed when $p$ approaching 715.6 nm, which is a typical signature of the formation of a BIC resonance [13]. To confirm this phenomenon is a consequence of total suppression of radiation loss, we performed an eigenfrequency analysis to calculate the radiative $Q$-factor of the studied resonance as a function of $p$. As plotted in Fig. 1(f), the $Q$-factor diverges near $p$ = 715.6 nm, indicating the resonance becomes a radiation-less dark states.

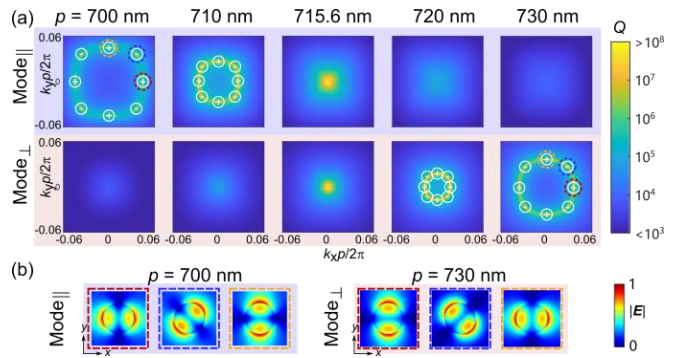

Fig. 2 (a) $Q$-factor mappings of Mode$_\parallel$ (upper) and Mode$_\perp$ (lower) in momentum space with varying lattice constant $p$. Eight accidental BICs with topological charge ±1 are labeled in the map. (b) Electric field profiles of accidental BICs formed in Mode$_\parallel$ and Mode$_\perp$ when $p$ = 700 nm and 730 nm.

To reveal the underlying physic in the formation of BIC in the lattice-constant-tuned resonant metasurface, we investigate the resonance behavior of Mode$_\parallel$ and Mode$_\perp$ in the momentum space. Figure 2 shows the 2D maps of the *Q*-factor of two modes with continuously varying $k_x$ and $k_y$ through an eigenfrequency calculation. When *p* = 700 nm, the *Q*-factor of Mode$_\parallel$ exhibit diverging behavior at eight points in the *k*-space, as depicted in Fig. 2(a), indicating the formation of accidental BICs at these off-Γ points [12]. Here $k_0$ is defined as $2\pi/p$. Each accidental BICs formed at off-Γ points carry a positive or negative topological charge (see Supplement 1 for detail), which are labelled in Fig.2(a). While Mode$_\perp$ cannot be transformed to accidental BICs in any point in the *k*-space, the *Q*-factor of which shows a monotonously decreasing trend with larger *k* values, as depicted in Fig. 2(a). The electric field profiles |*E*| of three typical accidental BICs circled in Fig. 2(a) at the cut plane at the quarter-height of the dielectric pillar of the Mode$_\parallel$ is illustrated in Fig. 2(b). When *p* = 710 nm, the positions of the accidental BICs of Mode$_\parallel$ in *k*-space becomes closer to Γ point, as depicted in Fig. 3(a). Mode$_\perp$ still does not show any sign of the formation of accidental BIC evidenced by diverging *Q*-factors. When *p* = 715.6 nm, the *Q*-factor of Mode$_\parallel$ diverges at Γ point, as a result of the merging of the eight accidental BICs at Γ point [20–23], and Mode$_\perp$ also shows a diverging *Q*-factor, due to the inherent degeneracy of Mode$_\parallel$ and Mode$_\perp$ at Γ point, which explains the vanishing linewidth of the transmission spectra in Fig. 2(a). When *p* is further increased to 730 nm, accidental BICs no longer emerges within Mode$_\parallel$, while Mode$_\perp$ have eight accidental BICs in momentum space, as shown in Fig. 2(a). It is worth noting that the topological charge of the degenerate merging BIC is zero, which differs from the case of merging BICs in non-degenerate modes [20,21].

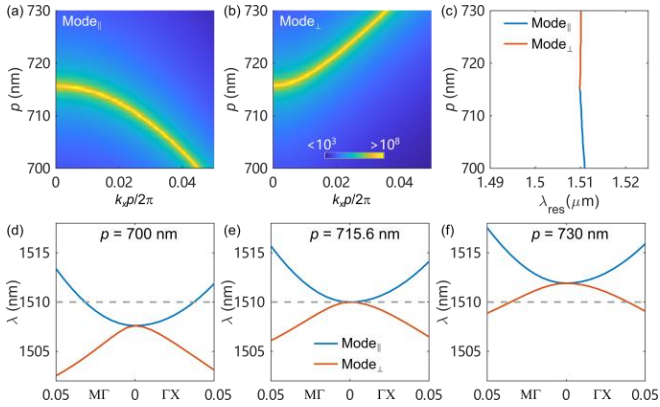

Fig. 3 *Q*-factor maps of (a) Mode$_\parallel$ and (b) Mode$_\perp$ in $k_x$-*p* space indicating the moving of accidental BICs along ΓX direction with continuously varying *p*. (c) The resonant wavelengths of accidental BICs in Mode$_\parallel$ and Mode$_\perp$ with different *p*. Band structure of Mode$_\parallel$ and Mode$_\perp$ with lattice constant equaling to (d) 700 nm, (e) 715.6 nm, and (f) 730 nm. Dashed line indicates the resonant wavelengths of BICs.

For a better illustration of the BIC merging process under continuous lattice constant changing, Figure 3(a-b) show the *Q*-factor mapping of Mode$_\parallel$ and Mode$_\perp$ in the $k_x$-*p* space, where we maintain $k_y$ = 0, and only consider the *p*-mediated accidental BIC shift along ΓX directions. For Mode$_\parallel$, with *p* increasing from 700 nm to 715.6 nm, diverging *Q*-factor in each *p* value indicates the accidental BIC continuously move towards at a lower $k_x$ value, while when *p* further increases to 730 nm, accidental BIC could no longer be formed, as depicted in Fig. 3(a). For Mode$_\perp$, accidental BIC manifested by the diverging *Q* only occurs when *p* is larger than 715.6 nm, and would move towards higher $k_x$ values with growing *p*. For both modes, *Q*-factors diverge simultaneously at $k_x$ = 0 (i.e. Γ point) with *p* = 715.6 nm, as a result of merging BICs. The BIC merging process along other directions are similar.

Another question may rise: why the accidental BICs in Mode$_\parallel$ and Mode$_\perp$ behaves so differently upon lattice constant changing. Recently, it is revealed that the accidental BICs can be viewed as a Friedrich-Wintgen BIC (FW-BIC) arising from the interaction between GMR and Fabry-Perot (FP) resonance [29]. When we treat the resonant metasurface as a homogeneous dielectric slab with the same thickness and its refractive index can be estimated from effective medium theory, FP modes residing in such dielectric slab could couple with the GR mode (see Supplement 1 for detail). In our case, the filling ratio (FR) of the dielectric pillar upon the changing of lattice constant nearly remain the same (i.e. FR is 25.6% and 23.6% with *p* equaling to 700 nm and 730 nm, respectively), leading to a nearly constant effective index of the equivalent homogeneous dielectric slab and thus a nearly constant resonant wavelength of FP mode. While for the GMR mode, its resonant wavelength is sensitive to the change of *p*. The resonant wavelengths of Mode$_\parallel$ and Mode$_\perp$, which are similar to GMRs in the high contrast grating [30,31], would experience a red shift upon an enlarged *p*. Accidental BIC could be formed in the vicinity of the intersection point of the FP and GMR bands, where the radiative loss is suppressed by the destructive interference originated from Friedrich–Wintgen mechanism. To validate the underlying physics of accidental BIC formation, we calculate the resonant wavelength of accidental BICs in Mode$_\parallel$ and Mode$_\perp$ with different *p* values, as plotted in Fig. 3(c). The resonant wavelengths remain around 1510 nm for both modes during *p* changing from 700 nm to 730 nm, resulted from the constant resonant wavelength of FP mode. When the lattice constant is 700 nm, the FP mode resonant wavelength, as indicated by the grey dashed line in Fig. 3(d), crosses the upper branch (Mode$_\parallel$), leading to the formation of accidental BICs in Mode$_\parallel$. With the *p* further growing to 715.6 nm, the dispersion band would move to larger wavelengths, leading to the crossing of FP mode and GR mode closer to Γ point. At *p* = 715.6 nm, the band of FP mode crosses the touching point of two bands at Γ point, where Mode$_\parallel$ and Mode$_\perp$ are degenerate, leading to the merging of accidental BICs. Larger *p* would further shift two bands towards longer wavelengths, and FP mode band starts to have intersections with lower branch corresponding to Mode$_\perp$, which explains only Mode$_\perp$ has accidental BICs when *p* is larger than 715.6 nm.

Merging BICs at Γ point also features largely enhanced *Q*-factor. The *Q*-factor of single symmetry-protected BICs is shown to possessing the scaling law Q ∝ $k^{-2}$, while the merging BICs could follow a scaling law of Q ∝ $k^{-6}$ or even $k^{-8}$, depending on the number of merged accidental BICs in the vicinity of the Γ point [21,23]. In our case, the *Q*-factor at the merging BIC with *p* = 715.6 nm shows a dependence of $k^{-4}$, as depicted in Fig. 4(a). We also plot the *Q*-factor of a single accidental BIC at with *p* = 700 nm, which shows a scaling law of Q ∝ $k^{-2}$, in stark contrast to the merging BIC scenario. Here, for the better comparison, Δ*k* is defined as *k* - $k_{BIC}$, and $k_{BIC}$ equals to 0 and 0.0044$k_0$ for merging BIC and accidental BIC, respectively. The *Q*-factor of merging BIC is nearly two orders of magnitude higher than that of a single BIC near their designed *k* vectors, which is highly important for maintaining a high-quality resonance, especially for finite-sized resonant metasurfaces.

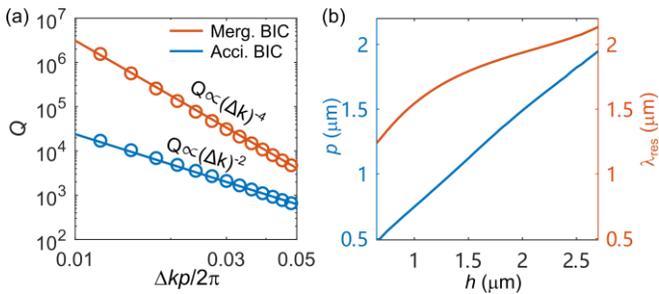

Fig. 4 (a) The scaling law of *Q*-factor of merging BIC and accidental BIC with respect to *k*. (b) The linear dependence of *P* on *h* where BIC occurs, and corresponding resonant wavelength is shown as red curve.

Tuning the BIC resonant wavelength is crucial for tailoring high-*Q* resonance response at targeted frequency ranges. As depicted in Fig. 4(b), when the height of pillar increases, the corresponding period where merging BIC happens shows a linear dependence on *h*. When *h* is 650 nm, the period becomes 469.3 nm to secure the resonant metasurface being in BIC state, and if *h* is further reduced, *p* will approach the diameter of the dielectric pillars (i.e. 400 nm), the resonance could no longer be perturbed into a radiative quasi-BIC state, which actually equals to the guided mode in a dielectric slab with the same height of pillar. We could observe that the resonant wavelength shows a red shift with a larger *h*, however, the wavelength is gradually approaching *p* with a larger *h*, and BIC state could no longer be sustained because the opening of high order diffraction channels, leading to extra radiation losses.

Note that degenerate merging BICs can provide much more fabrication tolerance on structural parameters (i.e., period, diameter) because there is always a resonance approaching to the merging BIC. Besides, for merging BICs in Ref. [20], since there is always a SP-BIC at Γ-point before and after merging process, making it impossible to be accessed from the farfield, which is not desired for some applications. For example, the merging BIC cavity for microlasers exhibit a donut shaped farfield radiation pattern with a large beam divergence [32], which is not suitable for directional lasing emission [33]. Thus, it is necessary to construct a merging BIC where BIC at Γ-point does not exist before and after merging process. Degenerate merging BICs at Γ-point may provide an alternative solution on this. In addition, the presence of a substrate could break the out-of-plane $\sigma_z$ symmetry, tuning the lattice constant could no longer achieve complete radiation suppression, and the highest *Q*-factor with a nonzero refractive index difference ($\Delta n$) between super- and substrate follows a scaling law as $Q \propto \Delta n^{-3}$ (see Supplement 1 for detail).

In summary, we show degenerate merging BICs could transform the radiative GMR resonance into a dark state in a resonant metasurface. After examining the lattice-constant-tuned resonance behavior in the momentum space, we found radiation suppression is originated from the merging of multiple accidental BICs at the Γ point. When an in-plane wavevector is presented, the guided mode resonance pair, i.e. Mode$_\parallel$ and Mode$_\perp$, become nondegenerate, and when $p < (>) p_c$, accidental BICs only occur in Mode$_\parallel$ (Mode$_\perp$) at off-Γ points. The nontrivial characteristics is because the coupling between FP and GR modes would lead to the formation of accidental BIC, and the GR modes is more sensitive to the changing of lattice constant, while the FP mode is more stable against the change, and thus the position of accidental BICs in the momentum space can be tuned through manipulating *p*. We also found the *Q*-factor of merging BIC shows a scaling law between *Q*-factor and the in-plane wavevector as $Q \propto \Delta k^{-4}$, indicating greatly increased *Q*-factors of the nearby modes in the same band. Our findings would provide a new insight for the merging BICs and would offer a simple route for constructing resonant metasurfaces with robust high-*Q* performance.

**Funding.** National Natural Science Foundation of China (62105172), Shanghai Pujiang Program (22PJ1402900).

**Disclosures.** The authors declare no conflicts of interest.

**Data availability.** The data of this study are available from the corresponding authors upon reasonable request.

**Supplemental document**. See Supplement 1 for supporting content.

# Degenerate merging BICs in resonant metasurfaces: supplemental document

## 1. Theoretical calculation method

For the calculation of eigenmode in the unit cell, we use a commercial software COMSOL Multiphysics based on finite-element method to perform the numerically eigenfrequency analysis. Floquet periodic boundary conditions are imposed in the x-y plane, and perfectly matched layers are set along z-direction to absorb the radiation field, and the distance between micropillar and PMLs are at least two wavelengths, in order to avoid near field perturbations. The mesh size is λ/6, where λ is the wavelength in the medium. Without loss of generality, we consider the resonant metasurface has a refractive index of 3.5, which is embedded in a homogeneous medium with a refractive index of 1.

For the calculation of transmission spectra, we use a MATLAB-based rigorous coupled wave analysis (RCWA) software, RETICOLO to analyze the spectral behavior of the investigated resonant metasurface. We also use a parfor loop to reducing the calculation time.

## 2. Calculation of the topological charge of BICs

BICs in resonant metasurfaces are vortex centers in the polarization directions of farfield radiation, which carries quantized topological charges, [1,2] defined as

$$q = \frac{1}{2\pi} \oint_L d\mathbf{k} \cdot \nabla_k \phi(\mathbf{k}) \qquad (S1)$$

where $L$ is a closed path around a BIC in momentum-space and $\phi(\mathbf{k})$ is the azimuthal angle of the polarization states' major axis. The polarization states can be described by a vector $c = (c_x, c_y)$, where $c_x$ and $c_y$ are the projected $x$ and $y$ components of the electric field, and thus $\phi(\mathbf{k}) = \arg[c_x(\mathbf{k}) + ic_y(\mathbf{k})]$. We use an eigenfrequency solver (COMSOL Multiphysics) to calculate the eigenmodes in $k$-space. The projected polarization vector could be calculated as

$$\mathbf{c}(k_x, k_y) = (c_x, c_y, c_z) = \iint_{cell} e^{-ik_x x - ik_y y} \mathbf{E}(k_x, k_y) dxdy \Big/ \iint_{cell} dxdy \qquad (S2)$$

## 3. Interaction between Fabry-Perot (FP) mode and guided mode resonance (GMR)

In the main text, the merging BIC is achieved through tuning lattice constant $p$. To further illustrate the effect of FP mode in the process of merging BIC, we consider a similar case where pillar height $h$ is tuned while the lattice constant $p$ is fixed. Here, the pillar radius $r$ and $p$ of the resonant metasurface is fixed at $r = 200$ nm and $p = 710$ nm, in such case, we could have an effective refractive index of the resonant metasurface when treated as a homogenous slab: for the resonant metasurface composed of silicon pillars ($n_{pillar} = 3.5$) embedded in air, the effective refractive index $n_{eff} = ff \cdot n_{pillar} + (1 - ff) \cdot n_{air} = 1.62$, where filling factor is $ff = \pi r^2 / p^2$. In such case, we could also define a refractive contrast $\Delta n = n_{pillar} - n_{air} = 2.5$.

Next, we keep the $n_{avg}$ is 1.62 and vary $\Delta n$. When $\Delta n = 0$, the resonant metasurface becomes a dielectric slab. Figure S1(a) shows the transmission spectra of normally incident plane wave under varying $h$. A clear Fabry-Perot resonance could be observed, where free spectral range become smaller at larger $h$.

When $\Delta n = 1.25$, the slight index modulation introduces guided mode resonances, as could be observed from the additional transmission peaks in Fig. S1(b). The intersection between F-P resonances and GMRs could lead to a vanishing linewidth of GMR, for example, as denoted by the arrow.

When the index modulation $\Delta n$ becomes larger, the linewidth of GMRs becomes larger owing larger radiative losses, in the intersecting region between GMR and FP mode, the vanishing linewidth of transmission dip could also be observed. The region denoted by the dashed circle is near the case considered in main text.

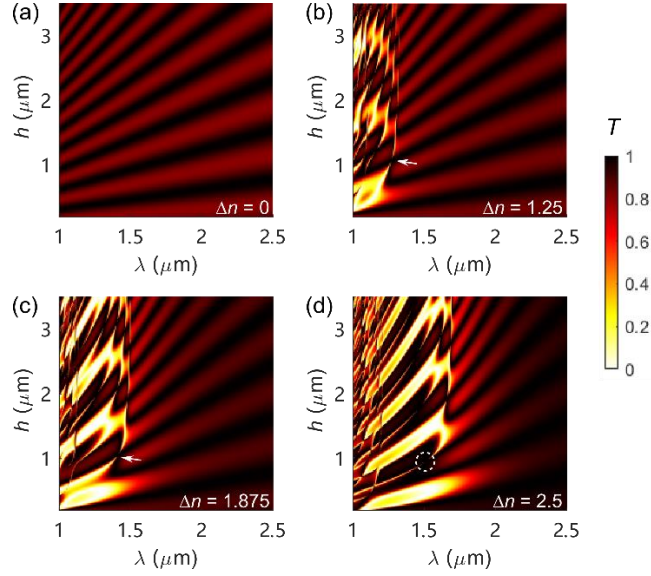

**Figure S1.** The coupling between FP mode and GMR mode. The transmission spectra under normally incident plane wave with different $h$ with (a) $\Delta n = 0$, where the resonant metasurface becomes a homogeneous slab. (b) $\Delta n = 1.25$. (c) $\Delta n = 1.875$. (d) $\Delta n = 2.5$. White arrows point out the exemplary positions where GMR linewidth vanishes. The region labeled by the dashed circle indicates the case considered in the main text. The other parameters are $d = 200$ nm, and $p = 710$ nm.

## 4. The influence of a substrate on the degenerate merging BICs

For practical realizations, the resonant metasurfaces are usually resting on a substrate for mechanical support. The presence of a substrate would break the out-of-plane mirror symmetry ($\sigma_z$) possessed by the metasurface embedded in a homogeneous medium. Figure S2(a) shows the transmission spectra of a resonant metasurface on a silica substrate with a refractive index of 1.44. The parameters of the pillars are the same as Fig. 2 of the main text. We could observe that a similar trend of the resonance linewidth evolution with varying $p$. However, the main difference is that the linewidth could not reach zero: the narrowest linewidth occurs at $p = 684.1$ nm, which corresponding to a $Q$-factor of $7.6 \times 10^4$. This phenomenon indicates a radiation-less BIC state could no longer be formed with a broken $\sigma_z$ symmetry. Figure S2(b) shows the variation of $Q$-factor of the resonance with the minimal linewidth under different substrate refractive indices. Here, refractive index asymmetry parameter $\Delta n$ is defined as $n_{sub} - 1$, because we consider air as the superstrate of the metasurface. The $Q$-factor follows an inverse cube law with $\Delta n$ as $1/(\Delta n)^3$. The underlying physics of the BIC collapse under out-of-plane permittivity asymmetry is that the at-$\Gamma$ BIC is originated from the merging of accidental BICs, while the accidental BIC is formed by destructive interferences which suppress radiation in both upward and downward out-of-plane directions. For resonant metasurface embedded in homogeneous medium, the position of accidental BIC in momentum space is dependent on the ambient refractive index. When the refractive index of the substrate and superstrate is not the same, the suppressed radiation to the upper and lower semi-spaces would occur at different in-plane wavevectors, leading to the collapse of accidental BICs.

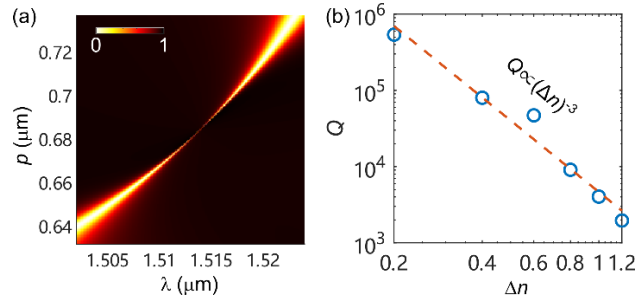

**Figure S2.** (a) Transmission spectra of a resonant metasurface resting on a silica substrate with varying $p$. (b) The scaling law of $Q$-factor of the merging BIC resonance with the presence of a substrate. $\Delta n$ characterizes refractive index asymmetry which is defined as $n_{sub} - 1$.

## 5. Comparison of merging BICs in various metasurfaces

Table R1. Comparison of merging BIC in various lattice configurations

| Ref. | Lattice | Before merging | | Merging | |
|---|---|---|---|---|---|
| | | T.C. | # of a-BICs | Merging position | Scaling law of m-BIC |
| [2,3] | triangular | -2 | 12 | At $\Gamma$ | $Q \sim k^{-8}$ |
| [4] | square | +1 | 8 | At $\Gamma$ | $Q \sim k^{-6}$ |
| [5] | square | +1 | 4 | At $\Gamma$ | $Q \sim k^{-6}$ |
| [5,6] | square | 0 | 2 | Off $\Gamma$ | $Q \sim (k - k_{BIC})^{-4}$ |
| This work | square | 0 | 8 | At $\Gamma$ | $Q \sim k^{-4}$ |